\count100=1 
\def\doit#1#2{\ifcase#1\or#2\fi}

\font\midlarge=cmbx10 scaled \magstep 2
\font\LARGE=cmr10 scaled \magstep 3

\catcode`@=11 \catcode`@=12

\doit1{ 
\font\eightrm=cmr8 
\font\eightbf=cmbx8 
\font\eightit=cmti9 
\font\eightsl=cmss8 
\font\eightmus=cmmi8 
\def\smalltype{\let\rm=\eightrm \let\bf=\eightbf \let\it=\eightit
\let\sl=\eightsl \let\mus\eightmus 
\baselineskip=1.5pt minus .75pt\rm} 
\def\it{\fam\itfam\tenit} 
} 

\def\footnotes#1#2{\footnote{$\,^{\star#1)}$}{\smalltype #2}} 
\def\newpage{\vfill\eject}

\def\today{\ifcase\month\or January\or February\or March\or
April\or May\or June\or July\or August\or September\or 
October\or November\or December\fi \space\number\day, 
\number\year} 

\let\du=\d 

\def\a{\alpha} \def\b{\beta}  \def\d{\delta}
\def\e{\epsilon}  \def\g{\gamma}
   
\def\l{\lambda} \def\m{\mu}  \def\o{\omega}
   \def\s{\sigma}
   \def\z{\zeta}
  \def\G{\Gamma}

\def\pmb#1{\setbox0=\hbox{${#1}$}%
   \kern-.025em\copy0\kern-\wd0
   \kern-.035em\copy0\kern-\wd0
   \kern.05em\copy0\kern-\wd0
   \kern-.035em\copy0\kern-\wd0
   \kern-.025em\box0 }


\def\bo{{\raise-.46ex\hbox{\large$\Box$}}} 

\def\pr{\prod} 

\def\TH{{\raise.2ex\hbox{$\displaystyle \bigodot$}\mskip-4.7mu %
\llap H \;}}
\def\face{{\raise.2ex\hbox{$\displaystyle \bigodot$}\mskip-2.2mu %
\llap {$\ddot
    \smile$}}} 

\def\sp#1{{}^{#1}} 
\def\sl#1{\rlap{\hbox{$\mskip 1 mu /$}}#1} 

 %
 %

\def\Tilde#1{{\widetilde{#1}}\hskip 0.015in} 
\def\Bar#1{\overline{#1}} 
\def\leftrightarrowfill{$\mathsurround=0pt \mathord\leftarrow
 \mkern-6mu
    \cleaders\hbox{$\mkern-2mu \mathord- \mkern-2mu$}\hfill
    \mkern-6mu \mathord\rightarrow$}
\def\dvec#1{\vbox{\ialign{##\crcr
    \leftrightarrowfill\crcr\noalign{\kern-1pt\nointerlineskip}
    $\hfil\displaystyle{#1}\hfil$\crcr}}} 

\def\frac#1#2{{\textstyle{#1\over\vphantom2\smash{\raise.20ex
    \hbox{$\scriptstyle{#2}$}}}}} 
\def\sfrac#1#2{{\vphantom1\smash{\lower.5ex\hbox{\small$#1$}}%
\over\vphantom1\smash{\raise.4ex\hbox{\small$#2$}}}}
\def\bfrac#1#2{{\vphantom1\smash{\lower.5ex\hbox{$#1$}}\over
    \vphantom1\smash{\raise.3ex\hbox{$#2$}}}} 
\def\afrac#1#2{{\vphantom1\smash{\lower.5ex\hbox{$#1$}}\over#2}} 


\newskip\humongous \humongous=0pt plus 1000pt minus 1000pt
\def\caja{\mathsurround=0pt}

\newif\ifdtup
\def\panorama{\global\dtuptrue \openup2\jot \caja
    \everycr{\noalign{\ifdtup \global\dtupfalse
    \vskip-\lineskiplimit \vskip\normallineskiplimit
    \else \penalty\interdisplaylinepenalty \fi}}}
\def\li#1{\panorama \tabskip=\humongous 
    \halign to\displaywidth{\hfil$\displaystyle{##}$
    \tabskip=0pt&$\displaystyle{{}##}$\hfil
    \tabskip=\humongous&\llap{$##$}\tabskip=0pt
    \crcr#1\crcr}}

\doit0{
\def\ref#1{$\sp{#1)}$}
}

\hsize=6in 
\parskip=\medskipamount 
\lineskip=0pt 
\abovedisplayskip=1em plus.3em minus.5em 
\belowdisplayskip=1em plus.3em minus.5em 
\abovedisplayshortskip=.5em plus.2em minus.4em 
\belowdisplayshortskip=.5em plus.2em minus.4em 

\def\endtitle{\end{quotation}\newpage} 

\def\sect#1{\bigskip\medskip \goodbreak \noindent{\bf {#1}} %
\nobreak \medskip}
\def\refs{\sect{References} \footnotesize \frenchspacing \parskip=0pt}
\def\hang{\hangindent\parindent}
\def\textindent#1{\indent\llap{#1\enspace}\ignorespaces} 
\def\Item{\par\hang\textindent}

\def\[{\lfloor{\hskip 0.35pt}\!\!\!\lceil}
\def\]{\rfloor{\hskip 0.35pt}\!\!\!\rceil}
\def\delsl{{{\partial\!\!\! /}}}
\def\nablasl{{{\nabla\!\!\!\!\!{\hskip 1.0pt}/}}}

\def\calO{{\cal O}}

\def\Lag{{\cal L}} 
\def\du#1#2{_{#1}{}^{#2}}
\def\ud#1#2{^{#1}{}_{#2}}

\def\calM{{\cal M}}\def\calO{{\cal O}}

\def\rma{{\rm a}} \def\rmb{{\rm b}} \def\rmc{{\rm c}}
\def\rmd{{\rm d}}
\def\rme{{\rm e}} \def\rmf{{\rm f}} \def\rmg{{\rm g}}
\def\rmh{{\rm h}} \def\rmi{{\rm i}} \def\rmj{{\rm j}}

\def\plpl{{+\!\!\!\!\!{\hskip 0.009in}%
{\raise-1.0pt\hbox{$_+$}} {\hskip 0.0008in}}}
\def\mimi{{-\!\!\!\!\!{\hskip 0.009in}%
{\raise-1.0pt\hbox{$_-$}} {\hskip 0.0008in}}}

\def\E{{\cal E}}

\def\pl#1#2#3{Phys.~Lett.~{\bf {#1}B} (19{#2}) #3}
\def\np#1#2#3{Nucl.~Phys.~{\bf B{#1}} (19{#2}) #3}

\def\pr#1#2#3{Phys.~Rev.~{\bf D{#1}} (19{#2}) #3}

\def\jhep#1#2#3{Jour.~High Energy Phys.~{\bf {#1}} (19{#2}) #3}

\def\mpl#1#2#3{Mod.~Phys.~Lett.~{\bf A{#1}} (19{#2}) #3}

\font\texttt=cmr10 
\def\hepth#1{{\texttt hep-th/{#1}}}

\def\npn#1#2#3{Nucl.~Phys.~{\bf B{#1}} (20{#2}) #3}

\def\jhepn#1#2#3{Jour.~High Energy Phys.~{\bf {#1}} (20{#2}) #3}

\def\<<{<\!\!<} \def\>>{>\!\!>}
\def\Check#1{{\raise-1.0pt\hbox{\LARGE\v{}}{\hskip -10pt}{#1}}}

\def\eqques{{~\,={\hskip -11.5pt}\raise -1.8pt\hbox{\large ?}
{\hskip 4.5pt}}{}}
\def\fracm#1#2{\,\hbox{\large{${\frac{{#1}}{{#2}}}$}}\,}
\def\fracmm#1#2{\,{{#1}\over{#2}}\,}

\def\frac#1#2{{\textstyle{#1\over\vphantom2\smash{\raise -.20ex
    \hbox{$\scriptstyle{#2}$}}}}} 

\def\scst{\scriptstyle}

\def\.{.$\,$}
\def\-{{\hskip 1.5pt}\hbox{-}}

\def\low#1{\hskip0.01in{\raise -2.3pt\hbox{${\hskip 1.0pt}\!_{#1}$}}}
\def\Low#1{\hskip0.015in{\raise -3pt\hbox{$\!_{#1}$}}}
\def\lowlow#1{\hskip0.01in{\raise -7pt%
\hbox{${\hskip1.0pt} \hskip-2pt_{#1}$}}}
\def\ldu#1#2{\low{#1}{}^{#2}}

\def\tr{\,{\rm tr}\,}

\font\tenmib=cmmib10
\font\sevenmib=cmmib10 at 7pt 
\font\fivemib=cmmib10 at 5pt 
\font\tenbsy=cmbsy10
\font\sevenbsy=cmbsy10 at 7pt 
\font\fivebsy=cmbsy10 at 5pt 
\def\BMfont{\textfont0\tenbf \scriptfont0\sevenbf
                  \scriptscriptfont0\fivebf
        \textfont1\tenmib \scriptfont1\sevenmib
                   \scriptscriptfont1\fivemib
        \textfont2\tenbsy \scriptfont2\sevenbsy
                   \scriptscriptfont2\fivebsy}
\def\rlx{\relax\leavevmode}
\def\BM#1{\rlx\ifmmode\mathchoice
              {\hbox{$\BMfont#1$}}
              {\hbox{$\BMfont#1$}}
              {\hbox{$\scriptstyle\BMfont#1$}}
              {\hbox{$\scriptscriptstyle\BMfont#1$}}
         \else{$\BMfont#1$}\fi}

\font\tenmib=cmmib10
\font\sevenmib=cmmib10 at 7pt 
\font\fivemib=cmmib10 at 5pt 
\font\tenbsy=cmbsy10
\font\sevenbsy=cmbsy10 at 7pt 
\font\fivebsy=cmbsy10 at 5pt 
\def\BMfont{\textfont0\tenbf \scriptfont0\sevenbf
                  \scriptscriptfont0\fivebf
        \textfont1\tenmib \scriptfont1\sevenmib
                   \scriptscriptfont1\fivemib
        \textfont2\tenbsy \scriptfont2\sevenbsy
                   \scriptscriptfont2\fivebsy}
\def\BM#1{\rlx\ifmmode\mathchoice
              {\hbox{$\BMfont#1$}}
              {\hbox{$\BMfont#1$}}
              {\hbox{$\scriptstyle\BMfont#1$}}
              {\hbox{$\scriptscriptstyle\BMfont#1$}}
         \else{$\BMfont#1$}\fi}

\def\inbar{\vrule height1.5ex width.4pt depth0pt}
\def\sinbar{\vrule height1ex width.35pt depth0pt}
\def\ssinbar{\vrule height.7ex width.3pt depth0pt}

\def\Ik{\rlx{\rm I\kern-.18em k}} 
\def\IC{\rlx\leavevmode
         \ifmmode\mathchoice
            {\hbox{\kern.33em\inbar\kern-.3em{\rm C}}}
            {\hbox{\kern.33em\inbar\kern-.3em{\rm C}}}
            {\hbox{\kern.28em\sinbar\kern-.25em{\rm C}}}
            {\hbox{\kern.25em\ssinbar\kern-.22em{\rm C}}}
         \else{\hbox{\kern.3em\inbar\kern-.3em{\rm C}}}\fi}
\def\IP{\rlx{\rm I\kern-.18em P}}
\def\IR{\rlx{\rm I\kern-.18em R}}
\def\IN{\rlx{\rm I\kern-.20em N}}
\def\Ione{\rlx{\rm 1\kern-2.7pt l}}

%

\newbox\leftpage \newdimen\fullhsize \newdimen\hstitle%
\newdimen\hsbody%
\tolerance=1000\hfuzz=2pt
\catcode`\@=11 
\hsbody=\hsize \hstitle=\hsize 

\def\nolabels{\def\wrlabeL##1{}\def\eqlabeL##1{}%
\def\reflabeL##1{}}
\def\writelabels{\def\wrlabeL##1{\leavevmode%
\vadjust{\rlap{\smash%
{\line{{\escapechar=` \hfill\rlap{\sevenrm\hskip.03in\string##1}}}}}}}%
\def\eqlabeL##1{{\escapechar-1%
\rlap{\sevenrm\hskip.05in\string##1}}}%
\def\reflabeL##1{\noexpand\llap{\noexpand%
\sevenrm\string\string%
\string##1}}} \nolabels
%
\global\newcount\secno \global\secno=0 \global\newcount\meqno
\global\meqno=1
\def\newsec#1{\global\advance\secno by1\message{(\the\secno. #1)}
\global\subsecno=0\eqnres@t\noindent{\bf\the\secno. #1}
\writetoca{{\secsym} {#1}}\par\nobreak\medskip\nobreak}
\def\eqnres@t{\xdef\secsym{\the\secno.}\global\meqno=1
\bigbreak\bigskip}
\def\sequentialequations{\def\eqnres@t{\bigbreak}}\xdef\secsym{}
\global\newcount\subsecno \global\subsecno=0
\def\subsec#1{\global\advance\subsecno by1%
\message{(\secsym\the\subsecno.%
 #1)}
\ifnum\lastpenalty>9000\else\bigbreak\fi
\noindent{\it\secsym\the\subsecno. #1}\writetoca{\string\quad
{\secsym\the\subsecno.} {#1}}\par\nobreak\medskip\nobreak}
\def\appendix#1#2{\global\meqno=1\global\subsecno=0%
\xdef\secsym{\hbox{#1.}} \bigbreak\bigskip\noindent{\bf Appendix
#1. #2}\message{(#1. #2)} \writetoca{Appendix {#1.}
{#2}}\par\nobreak\medskip\nobreak}
\def\eqnn#1{\xdef #1{(\secsym\the\meqno)}%
\writedef{#1\leftbracket#1}%
\global\advance\meqno by1\wrlabeL#1}
\def\eqna#1{\xdef #1##1{\hbox{$(\secsym\the\meqno##1)$}}
\writedef{#1\numbersign1\leftbracket#1{\numbersign1}}%
\global\advance\meqno by1\wrlabeL{#1$\{\}$}}
\def\eqn#1#2{\xdef #1{(\secsym\the\meqno)}%
\writedef{#1\leftbracket#1}%
\global\advance\meqno by1$$#2\eqno#1\eqlabeL#1$$}
%
\newskip\footskip\footskip8pt plus 1pt minus 1pt
\font\smallcmr=cmr5
\def\footnotefont{\smallcmr}
\def\f@t#1{\footnotefont #1\@foot}
\def\f@@t{\baselineskip\footskip\bgroup\footnotefont\aftergroup%
\@foot\let\next}
\setbox\strutbox=\hbox{\vrule height9.5pt depth4.5pt width0pt} %
\global\newcount\ftno \global\ftno=0
\def\foot{\global\advance\ftno by1\footnote{$^{\the\ftno}$}}
%
\newwrite\ftfile
\def\footend{\def\foot{\global\advance\ftno by1\chardef\wfile=\ftfile
$^{\the\ftno}$\ifnum\ftno=1\immediate\openout\ftfile=foots.tmp\fi%
\immediate\write\ftfile{\noexpand\smallskip%
\noexpand\item{f\the\ftno:\ }\pctsign}\findarg}%
\def\footatend{\vfill\eject\immediate\closeout\ftfile{\parindent=20pt
\centerline{\bf Footnotes}\nobreak\bigskip\input foots.tmp }}}
\def\footatend{}
\global\newcount\refno \global\refno=1
\newwrite\rfile
\def\ref{[\the\refno]\nref}%
\def\nref#1{\xdef#1{[\the\refno]}\writedef{#1\leftbracket#1}%
\ifnum\refno=1\immediate\openout\rfile=refs.tmp\fi%
\global\advance\refno by1\chardef\wfile=\rfile\immediate%
\write\rfile{\noexpand
\Item{#1}\reflabeL{#1\hskip-10pt}\pctsign}%
\findarg\hskip10.0pt}%
\def\findarg#1#{\begingroup\obeylines\newlinechar=`\^^M\pass@rg}
{\obeylines\gdef\pass@rg#1{\writ@line\relax #1^^M\hbox{}^^M}%
\gdef\writ@line#1^^M{\expandafter\toks0%
\expandafter{\striprel@x #1}%
\edef\next{\the\toks0}\ifx\next\em@rk\let\next=\endgroup%
\else\ifx\next\empty%
\else\immediate\write\wfile{\the\toks0}%
\fi\let\next=\writ@line\fi\next\relax}}
\def\striprel@x#1{} \def\em@rk{\hbox{}}
\def\lref{\begingroup\obeylines\lr@f}
\def\lr@f#1#2{\gdef#1{\ref#1{#2}}\endgroup\unskip}

\def\addref#1{\immediate\write\rfile{\noexpand\item{}#1}} 
%

%


%
\def\startrefs#1{\immediate\openout\rfile=refs.tmp\refno=#1}
\def\xref{\expandafter\xr@f}\def\xr@f[#1]{#1}
\def\refs#1{\count255=1[\r@fs #1{\hbox{}}]}
\def\r@fs#1{\ifx\und@fined#1\message{reflabel %
\string#1 is undefined.}%
\nref#1{need to supply reference \string#1.}\fi%
\vphantom{\hphantom{#1}}\edef\next{#1}%
\ifx\next\em@rk\def\next{}%
\else\ifx\next#1\ifodd\count255\relax\xref#1\count255=0\fi%
\else#1\count255=1\fi\let\next=\r@fs\fi\next}

\newwrite\ffile\global\newcount\figno \global\figno=1
\doit0{
\def\fig{fig.~\the\figno\nfig}
\def\nfig#1{\xdef#1{fig.~\the\figno}%
\writedef{#1\leftbracket fig.\noexpand~\the\figno}%
\ifnum\figno=1\immediate\openout\ffile=figs.tmp%
\fi\chardef\wfile=\ffile%
\immediate\write\ffile{\noexpand\medskip\noexpand%
\item{Fig.\ \the\figno. }
\reflabeL{#1\hskip.55in}\pctsign}\global\advance\figno
by1\findarg}
\def\xfig{\expandafter\xf@g}\def\xf@g fig.\penalty\@M\ {}
\def\figs#1{figs.~\f@gs #1{\hbox{}}}
\def\f@gs#1{\edef\next{#1}\ifx\next\em@rk\def\next{}\else
\ifx\next#1\xfig #1\else#1\fi\let\next=\f@gs\fi\next}
}

\newwrite\lfile
{\escapechar-1\xdef\pctsign{\string\%}\xdef\leftbracket{\string\{}
\xdef\rightbracket{\string\}}\xdef\numbersign{\string\#}}

\def\writestop{\def\writestoppt%
{\immediate\write\lfile{\string\pageno%
\the\pageno\string\startrefs\leftbracket\the\refno\rightbracket%
\string\def\string\secsym\leftbracket\secsym\rightbracket%
\string\secno\the\secno\string\meqno\the\meqno}%
\immediate\closeout\lfile}}
\def\writestoppt{}\def\writedef#1{}
\def\seclab#1{\xdef #1{\the\secno}\writedef{#1\leftbracket#1}%
\wrlabeL{#1=#1}}
\def\subseclab#1{\xdef #1{\secsym\the\subsecno}%
\writedef{#1\leftbracket#1}\wrlabeL{#1=#1}}
\newwrite\tfile \def\writetoca#1{}
\def\leaderfill{\leaders\hbox to 1em{\hss.\hss}\hfill}
\def\writetoc{\immediate\openout\tfile=toc.tmp
   \def\writetoca##1{{\edef\next{\write\tfile{\noindent ##1
   \string\leaderfill {\noexpand\number\pageno} \par}}\next}}}

%
\catcode`\@=12 
%

\countdef\pageno=0 \pageno=1
\newtoks\headline \headline={\hfil}
\newtoks\footline
\footline={\tenrm\folio} 
\def\folio{\ifnum\pageno<0 \romannumeral-\pageno
\else{\hss\number\pageno\hss}\fi}

\def\advancepageno{\ifnum\pageno<0 \global\advance\pageno by -1
\else\global\advance\pageno by 1 \fi}
\newif\ifraggedbottom



\def\circle#1{$\bigcirc{\hskip-9pt}\raise-1pt\hbox{#1}$}

\def\eqdot{~{\buildrel{\hbox{\LARGE .}} \over =}~}

\def\eqques{~{\buildrel ? \over =}~}

\def\binomial#1#2{\left(\,{\buildrel
{\raise4pt\hbox{$\displaystyle{#1}$}}\over
{\raise-6pt\hbox{$\displaystyle{#2}$}}}\,\right)}

\font\smallcmr=cmr6 scaled \magstep2 
scaled \magstep 1

\font\large=cmr10 scaled \magstep3




\font\smallcmr=cmr6 scaled \magstep2

\def\fracmm#1#2{{{#1}\over{#2}}}

\topskip 0.2in
 
\baselineskip 20.0pt
\vsize=9.0in
\hsize=6.5in
\pagedepth=-5in

\tolerance=10000
\magnification=1200

\pageno=1

\doit0{
{\bf Preliminary Version (FOR YOUR EYES ONLY!) \hfill \today}
} 
\vskip -0.03in

\doit1{
\rightline{hep-th/0502089} 
\vskip -0.1in 
}
\rightline{CSULB--PA--04--7} 
\vskip -0.0in 

\vskip 0.5in

\centerline{\midlarge Dual Vector Multiplet} 
\vskip 0.02in 
\centerline{\bf \midlarge Coupled to Dual $\,$N=1$\,$
Supergravity in 10D}

\bigskip

\baselineskip 9pt

\vskip 0.18in

\centerline{Hitoshi ~N{\smallcmr ISHINO}%
\footnotes1{E-Mail: hnishino@csulb.edu}
~and ~Subhash ~R{\smallcmr AJPOOT}%
\footnotes2{E-Mail: rajpoot@csulb.edu} 
}

\bigskip
\centerline{\it Department of Physics \& Astronomy}
\vskip 0.03in
\centerline{\it California State University} 
\vskip 0.03in
\centerline{\it 1250 Bellflower Boulevard} 
\vskip 0.03in
\centerline{\it Long Beach, CA 90840} 
\bigskip

\vskip 1.0 in

\centerline{\bf Abstract}
\bigskip

\vskip 0.1in

\baselineskip 14pt 

~~~We couple in superspace a `dual' vector multiplet 
$~(C_{m_1\cdots m_7}, \l^\a)$~ to the dual version of $~N=1$~
supergravity $~(e\du m a, \psi\du m\a,  M_{m_1\cdots m_6} ,
\chi\low\a,\Phi)$~  in ten-dimensions.  Our new 7-form field $~C$~ has
its 8-form field strength $~H$~ dual to the 2-form field strength $~F$~ of
the conventional vector multiplet.  We have found that the
$~H\-$Bianchi identity must have the form $~N\wedge F$, where
$~N$~ is the 7-form field strength in dual supergravity.  We also see  why
only the dual version of  supergravity couples to the dual vector 
multiplet consistently.  The potential anomaly for the 
dual vector multiplet can be cancelled for the particular gauge group
$~U(1)^{496}$~ by the Green-Schwarz mechanism.  As a by-product, we
also give the globally supersymmetric Abelian Dirac-Born-Infeld
interactions for the dual vector multiplet for the first time.

\vskip 0.35in 
\leftline{\smalltype  PACS:  ~04.65.+e, 11.30.Pb, 12.60.Jv, 
11.10.Lm, 11.15.Ðq} 
\vskip -0.05in 
\leftline{\smalltype  Key Words:  Supergravity, 
Supersymmetry, Superspace, Anomaly, Hodge Duality,
Ten-Dimensions} 
\vfill\eject

\baselineskip 18.0pt

\pageno=2
\leftline{\bf 1.~~Introduction}  

It is common wisdom in ten-dimensions (10D)  that only the vector
multiplet (VM) with a real vector 
$~A_m$\footnotes3{We use $~{\scst m,~n,~\cdots}$~ for curved
coordinates in this paper.} and a Majorana spinor $~\l$~ 
\ref\gso{F.~Gliozzi, J.~Scherk and D.~Olive, \np{122}{77}{253}; 
L.~Brink, J.H.~Schwarz and J.~Scherk, \np{121}{77}{77}.}    
have interactions consistent with supersymmetry.   For example, even
though the tensor field $~C_{m_1\cdots m_7}$, which is Hodge dual to
$~A_m$, has the same eight degrees  of freedom as the latter, there
have been certain  long-standing problems   with constructing
consistent interactions.  For example, non-Abelian  interactions with
such a high-rank tensor have been known to be  problematic. 
Additionally, this wisdom draws support from superstring theory
\ref\gsw{M.B.~Green, J.H.~Schwarz and E.~Witten, 
{\it `Superstring Theory'}, Vols.~I \& II.},  
because it is the vector field $~A_m$~ instead of 7-form tensor 
$~C_{\m_1\cdots\m_7}$~  that couples consistently to open
or heterotic strings \gso%
\ref\superstring{E.S.~Fradkin and A.A.~Tseytlin, 
\pl{163}{85}{123};  
A.~Abouelsaood, C.~Callan, C.~Nappi and S.~Yost, 
\np{280}{87}{599}.}%
\gsw.  

In component formulation in 10D \gso,  
the main technical obstruction for the `dual' VM 
multiplet $~(C_{m_1\cdots  m_7}, \l)$~ has been 
the necessity of an extra symmetry for the 
closure of supersymmetries ~$\[\d_Q(\e_1), 
\d_Q(\e_2)\]$, in addition to the usual global Poincar\'e and 
$~U(1)$~ gauge symmetries.  This seems to be a
persistent problem, because such an extra symmetry 
does not seem to be maintained in the presence of  
interactions.  In superspace language 
\ref\ggrs{S.J.~Gates, Jr., M.T.~Grisaru, M.~Ro\v cek 
and W.~Siegel, {\it `Superspace'}  (Benjamin/Cummings,
Reading, MA 1983);  
J.~Wess and J.~Bagger, {\it `Superspace and Supergravity'}, 
Princeton University Press (1992).}, 
this corresponds to the breakdown of $~H\-$Bianchi identities (BIs).  
Therefore, our two conventional tools for supersymmetry, {\it i.e.,}
component \gso\ and superspace \ggrs\ formulations seem to be of
limited use to solve the problem.  Naturally, so far very few papers have
been  devoted to solve this long-standing problem.     

In this paper, we give the first solution to this long-standing problem, by
coupling the dual VM $(C_{m_1\cdots m_7}, \l^\a)$~ to $~N=1$~ dual  
supergravity in 10D with the field content $~(e\du m a, \psi\du m \a, 
M_{m_1\cdots m_6}, \chi\low\a, \Phi)$~ 
\ref\chamseddinedual{A.~Chamseddine, \pr{24}{81}{3065}.}%
\ref\bdrdw{E.~Bergshoeff, M.~de Roo and B.~de Wit, 
\np{217}{83}{489}.}%
\ref\gnanomaly{J.S.~Gates, Jr.~and H.~Nishino, 
\pl{157}{85}{157}; \pl{173}{86}{52}; \np{291}{87}{205}.}%
\ref\gngsstring{S.J.~Gates, Jr.~and H.~Nishino, 
\pl{173}{86}{46}.}.       
We will show that the $~H\-$BI  for the superfield strength $~H=d C $~ is
modified by a term $~N\wedge F$~ 
composed of the superfield strength $~N = dM~$ in the dual
supergravity multiplet \gnanomaly\gngsstring\ and the
$~F\-$superfield strength of  the conventional vector multiplet.  

As a technical tool, we use the so-called 
`beta-function-favored-constraints' (BFFC) for the dual $~N=1$~ 
supergravity in 10D.  This is a very special set of constraints 
developed in ref.~%
\ref\bffcdual{H.~Nishino, \pl{258}{91}{104}.}.  
Originally, the BFFC for usual version of $~N=1$~ supergravity in 10D 
\ref\chamseddineusual{A.~Chamseddine, \np{185}{81}{403};
E.~Bergshoeff, M.~de Roo, B.~de Wit and P.~van Nieuwenhuizen,
\np{195}{82}{97};  G.F.~Chapline and N.S.~Manton, \pl{120}{83}{105}.} 
was developed in 
\ref\gnz{M.T.~Grisaru, H.~Nishino and D.~Zanon, 
\pl{206}{88}{605}, \np{314}{89}{363}.},  
in order to simplify $~\beta\-$function computations for the 
Green-Schwarz $~\s\-$model, by simplifying the constraint system.  
It was found \bffcdual\ that the BFFC set for the dual version of $~N=1$~ 
supergravity \gnanomaly\gngsstring\ has one of 
the simplest coupling structures for our dual VM in 10D.   

We also study possible anomalies for the Abelian dual VM.  Due to the
Abelian nature, there can only be the gravitational
anomalies.  Moreover, we find that the gravitational anomalies can be 
cancelled for the gauge group $~U(1)^{496}$~ by the Green-Schwarz 
mechanism 
\ref\gs{M.~Green and J.H.~Schwarz, \pl{149}{84}{117}; 
M.~Green, J.H.~Schwarz and P.~West, \np{254}{85}{327}; 
A.~Salam and E.~Sezgin, Physica Scripta {\bf 32} (1985) 283.}.   

In a sense, our approach here is similar to the claim about the  
couplings of vector multiplet to supergravity first presented by 
\chamseddinedual.  In that paper, it was claimed that 
a vector multiplet can be coupled to supergravity only {\it via} 
dual formulation.  However, this was pointed out to be 
incorrect in \bdrdw, because the duality transformation 
\ref\nt{H.~Nicolai and P.K.~Townsend, \pl{98}{81}{257}.}   
can connect the dual formulation with the usual formulation.  
However, the difference in our formulation seems to be that 
in the case of dual vector multiplet, due to its index structure, 
only the dual formulation of $~N=1$~ supergravity 
can be coupled to dual vector multiplet.  This point will be clarified 
in section 4.  

As a by-product, we also show that an Abelian dual VM can have 
supersymmetric Dirac-Born-Infeld (DBI) interactions
\ref\dbi{M.~Born and L.~Infeld, Proc.~Roy.~Soc.~Lond.~%
{\bf A143} (1934) 410; {\it ibid.}~{\bf A144} (1934) 425;
P.A.M.~Dirac, Proc.~Roy.~Soc.~Lond.~{\bf A268} (1962) 57.}%
\ref\uniquedbi{A.A. Tseytlin, \np{501}{97}{41}, \hepth{9701125};  
{\it `Born-Infeld Action, Supersymmetry \& String Theory'}, 
in the Yuri Golfand Memorial Volume, ed.~M.~Shifman, 
World Scientic (2000), \hepth{9908105}, {\it and references therein};
L.De Fosse, P.~Koerber and A.~Sevrin, 
\npn{603}{01}{413}, \hepth{0103015}; }%
\ref\cnt{M. Cederwall, B.E.W. Nilsson and D. Tsimpis, 
\jhepn{0106}{01}{034}, hep-th/0102009; 
\jhepn{0107}{01}{042}, \hepth{0104236}.}   
consistent with global supersymmetry.  We give the 
total lagrangian in component language 
at the order $\,{\rm (lengh)}^2\,$ with  
supersymmetry transformations.  This lagrangian is derived by 
duality transformation \nt\ 
from the conventional 2-form field strength $~F$~ into the 
dual 8-form strength $~H$.        

This paper is organized as follows:  In the next section, we first  clarify
the reason why we need the dual formulation 
\chamseddinedual\gnanomaly\bffcdual\ instead of  the usual
formulation \chamseddineusual\gnz\ of \hbox{$~N=1$}  supergravity in
10D for our purpose.     In section 3, we fist review the BFFC set for the
superspace formulation of dual $~N=1$~ supergravity in 10D
\bffcdual.   In section 4, we first review the coupling of the  dual BFFC
\bffcdual\ to the usual VM.  We next give our results of our dual VM
coupled to the dual $~N=1$~ supergravity \bffcdual.  In section 5, we
show that potential  anomaly can be cancelled for the gauge group
$~U(1)^{496}$, by  applying the Green-Schwarz mechanism \gs\ for the
dual formulation \gnanomaly\gngsstring.  In section 6, we give the
supersymmetric  Dirac-Born-infeld interaction with global
supersymmetry in  component language.  Some concluding remarks are
given in section 7.

\bigskip\bigskip\medskip 

\leftline{\bf 2.~~Necessity of Dual Supergravity} 

Before going into the detailed computations, we first clarify why only
the dual formulation of $~N=1$~ supergravity in 10D 
\chamseddinedual\gnanomaly\gngsstring\bffcdual\ is needed for our
purpose.   In other words, we clarify why we can not use
the usual formulation of supergravity \chamseddineusual\ which is
much simpler with fewer indices.  Such a question is motivated by the
expectation that a duality transformation \nt\ should be always
performed between the usual and dual formulations of supergravity.  

This question is also historically motivated.  In the paper
\chamseddinedual, where an $~N=1$~ non-Abelian VM was first coupled
to  supergravity in 10D, only  the dual formulation was claimed to be
coupled to a VM.  However, a later investigation \bdrdw\ showed  that
such a conclusion in \chamseddinedual\ was incorrect.   Ref.~\bdrdw\
showed that  a non-Abelian VM can be coupled not only to the dual
formulation of $~N=1$~ supergravity \chamseddineusual, but also to the
usual formulation \chamseddineusual.  The crucial reason is that a
duality transformation \nt\ can convert the dual formulation 
\chamseddinedual\ into  the usual formulation \chamseddineusual,
even in the presence of a VM.  

Based on such history, it is legitimate to expect a similar situation with
the coupling of a dual VM $~(C_{m_1\cdots m_7}, l)$,\footnotes4{We 
have to distinguish two different dualities here.  One between the usual
supergravity \chamseddineusual\ and  dual supergravity 
\chamseddinedual\gnanomaly\gngsstring, and another between the
VM $~{\scst (A_m,\,\l)}$~ and dual VM $~{\scst (C_{m_1\cdots m_7},\,
\l)}$.} {\it i.e.}, not only the dual \chamseddinedual\ but also the usual
\chamseddineusual\ formulation of supergravity can be used.  In this
section, we explain that this is not the case for coupling the dual VM.     

Consider the kinetic term for  the 2-form potential $~B_{m n}$~ in the
usual formulation \chamseddineusual\ of 
$~N=1$~ supergravity in 10D coupled to Abelian VM:\footnotes5{Our
metric is ${\scst (\eta_{a b} )  ~\equiv~\hbox{diag.}~(+, -, -, \cdots, -)}$. 
Relevantly, $~{\scst\e^{01\cdots 9} \equiv +1, ~
\g\low{11} \equiv \g\low{(0)(1)\cdots (9)}}$, where 
the indices $~{\scst a, ~b,~\cdots~ =~(0),  
~(1),~\cdots,~ (9)}$~ are for local Lorentz indices.  
We use  the antisymmetrization symbol normalized as 
$~{\scst A_{\[ m} B_{n\]} \equiv A_m B_n - A_n B_m}$~ {\it without} the
factor of 1/2 in front, due to  certain advantage in superspace
computation \ggrs.}     
$$ \li{ & -\frac1{12} e G_{m n r}^2 
      \equiv -\frac1{12} e \Big( \frac 12\partial _{\[m} B_{n r\]} 
      + \frac 12 F_{\[ m n} A_{r\]} \Big)^2 ~~. 
&(2.1) \cr } $$ 
Because of the `bare' potential field $~A$~ instead of its field strength 
$~F$~ in this Chern-Simons term in (2.1), we can not perform the duality
transformation \nt\ from the vector $~A_m$~ into its dual
$~C_{m_1\cdots m_7}$.  Notice also that the duality transformation  that
matters here is {\it not} between the dual version
\chamseddinedual\  and usual version \chamseddineusual\ of 
supergravity, but that between the VM and dual VM.  

On the contrary, in the dual formulation \chamseddinedual\ of 
supergravity, the corresponding coupling to (2.1) is put into the 
form 
$$\li{ & \e^{m_1\cdots m_6 n r s t} M_{m_1\cdots m_6} 
           F_{n r} F_{s t} ~~,    
&(2.2) \cr } $$   where only the field strength
$~F$~ with no `bare' potential $~A$~ is involved.  Therefore, the duality
transformation from the usual VM  into dual VM is possible.  In the usual
version \chamseddineusual, however, even if we try with partial
integrations {\it etc.}, we can not convert the Chern-Simons  term
$~F\wedge A$~ in (2.1) into $~F\wedge F$~ only with the  field strength
$~F$~ with no bare potential $~A$.  This obstruction  indicates that the
dual VM can not be coupled to the usual formulation
\chamseddineusual\ of supergravity.       

These points clarify why the dual formulation 
\chamseddinedual\ is needed for our purpose.  It is now clear that we 
use the dual formulation \chamseddinedual\gnanomaly\gngsstring\  as
the `necessity' for our couplings of the dual VM,  but not for simple
`curiosity'.

\bigskip\bigskip\medskip 


\leftline{\bf 3.~~Review of BFFC for Dual $~N=1$~ Supergravity}   

We next  review the main structure of BFFC superspace constraints
\bffcdual\  for the $~N=1$~ dual supergravity in 10D \gnanomaly, before
coupling our new dual VM to supergravity, .    The field content of the
dual $~N=1$~ supergravity is \bffcdual\ 
$~(e\du m a, \psi\du m \a,  M_{m_1\cdots m_6}, 
\chi\low\a, \Phi)$.   The supertorsion $~T\du{A B} C$, the supercurvature 
$~R\du{A B c} d$~ and, in particular, the 7-form superfield strength 
$~N_{A_1\cdots A_7}$~ of $~M_{A_1\cdots A_6}$~  characterize the
superspace formulation of the dual supergravity
\gnanomaly\gngsstring\bffcdual.  The usual VM has the field 
content $~(A_m{}^I, \l^{\a I})$~ with the superfield strength 
$~F_{A B}{}^I$, where $~^I$~ is the adjoint index for a non-Abelian 
gauge group.  

Accordingly, there are corresponding $~T$, $~N~$ and $~F\-$BIs  in
superspace\footnotes6{For the superspace 
 local Lorentz indices $~{\scst A~\equiv ~(a, \a), ~B~\equiv~(b, \b),
~\cdots}$, we use the indices $~{\scst a, ~b,~\cdots~ =~(0),
~(1),~\cdots,~ (9)}$~ for bosonic coordinates, while $~{\scst
\a,~\b,~\cdots ~=  ~1,~2,~\cdots,~16}$~ for chiral fermionic
coordinates.}  
$$ \li{ & \frac 12 \nabla_{\[A} T\du{B C)} D 
      - \frac 12 T\du{\[AB|} E  T\du{E|C)} D 
      - \frac 1 4 R\du{\[A B| e} f (\calM\du f e)\du{| C)} D \equiv 0 ~~,  
&(3.1\rma) \cr 
& \frac 1{7!} \nabla_{\[A_1} N_{A_2\cdots A_8)} 
     - \frac 1{6!\cdot 2} T\du{\[A_1 A_2 | } B N_{B | A_3\cdots A_8)} 
    \equiv 0 ~~,  
&(3.1\rmb) \cr 
& \frac1 2 \nabla_{\[A} F_{B C)}{}^I - \frac 12 T\du{\[A B |} D F_{D| C)}{}^I 
     \equiv 0 ~~. 
&(3.1\rmc) \cr  } $$ 

Our superspace  BFFC  for $~N=1$~ dual supergravity are 
\bffcdual\footnotes7{We use the symbols like $~{\scst 
\g^{\[3\]}}$, where generally $~{\scst \[ n\]}$~ stands 
for totally antisymmetric $~{\scst n}\,$-indices.}   
$$ \li{ & T\du{\a\b} c = + i(\g^c)_{\a\b} ~~, ~~~~ 
          N_{\a\b c_1\cdots c_5} = + \frac i 2 (\g_{c_1\cdots c_5} )_{\a\b} ~~, 
&(3.2\rma) \cr 
& T\du{\a b}\g = - \frac 1{36} (\g^{\[3\]} \g_b)\du \a\g \Tilde N_{\[3\]} 
     ~~, ~~~~ T\du{a b} c = + 2\Tilde N\du{a b} c ~~, 
&(3.2\rmb) \cr 
& \nabla_\a \chi\low\b = - \frac i{12\sqrt2} (\g^{\[3\]})_{\a\b} 
    \Tilde N_{\[3\]} - \frac i{\sqrt 2} (\g^c )_{\a\b} \nabla_c \Phi
     + \frac i {72\sqrt2} (\g^{\[3\]})_{\a\b} \chi\low{\[3\]} 
     + i (\g^{\[3\]})_{\a\b} A_{\[3\]} {\, , ~~~~~ ~~~~~}  
&(3.2\rmc) \cr 
& R\du{\a\b \, c} d = + \frac{5i}3 (\g^e)_{\a\b} \Tilde N\du{e c} d 
     + \frac i{18} (\g\du c{d \[3\]})_{\a\b} \Tilde N_{\[3\]} ~~, ~~~~
&(3.2\rmd) \cr 
& F_{\a b}{}^I = + \frac i{\sqrt2} (\g\low b \l{}^I )_\a ~~, 
&(3.2\rme) \cr 
& \nabla_\a\l^{\b I} = - \frac1{2\sqrt 2} (\g^{c d})\du\a\b F_{c d}{}^I ~~, 
&(3.2\rmf) \cr 
& R_{\a b c d} = - \frac i 2 (\g\low{\[c} T_{ d\] b} 
    - \g\low b T_{c d} )_\a + \frac i 4 (\g\low{b c d}{}^{e f} T_{e f} )_\a ~~,  
      ~~~~ \nabla_\a\Tilde N_{b c d}  
     = - \frac i 4 (\g\low{b c d}{}^{e f} T_{e f})_\a {~, ~~~~~ ~~~~~}  
&(3.2\rmg) \cr
& \nabla_\a F_{b c}{}^I  = - \frac i{\sqrt 2} 
      (\g\low{\[b} \nabla_{c\]} \l{}^I )_\a 
     - \frac {5i}{3\sqrt 2} (\g^d\l{}^I )_\a \Tilde N_{b c d}   \cr 
& ~~~~~ ~~~~~ ~~  + \frac i{6\sqrt 2} (\g\low{\[b | }{}^{\[2\]} \l{}^I )_\a 
          \Tilde N_{| c \] \[2\]} 
       - \frac i{18\sqrt2} (\g\low{b c}{}^{\[3\]} \l{}^I )_\a 
          \Tilde N_{\[3\]} ~~.   
&(3.2\rmh) \cr
& \nabla_\g T\du{a b}\d = - \frac 14(\g^{c d})\du\g\d 
      R_{a b c d} + \frac 1{36} (\g^{\[3\]} \g_{\[a|} )\du\g\d 
      \nabla_{ | b\]} \Tilde N_{c d e} \cr 
& ~~~~~ ~~~~~ ~ + \frac 1{18} (\g^{\[3\]} \g_e)\du\g\d 
     \Tilde N\du{a b} e \Tilde N_{\[3\]} 
     + \frac 1{1296} (\g^{\[3\]} \g_{\[a} \g^{\[3\]'} \g_{b\]} )\du\g\d
   \Tilde N_{\[3\]} \Tilde N_{\[3\]'} ~~, 
&(3.2\rmi) \cr 
& \Tilde N_{\[3\]} \equiv (1/7!) \e\low{\[3\]}{}^{\[7\]}  N\low{\[7\]}
     ~, ~~~ \chi\low{\[3\]} \equiv i (\Bar\chi\g\low{\[3\]} \chi) ~,     
     ~~~ A_{\[3\]} \equiv \frac i{32\sqrt2} \, e^{4\Phi/3}
      (\Bar\l{}^I \g\low{\[3\]} \l{}^I ) ~~, \hskip 0.26in
&(3.2\rmj) \cr } $$ 
As usual in superspace, all other independent components at 
$~d\le 1$, such as $~F_{\a\b}$~ or $~T\du{\a b} c$~ are zero.  
In order to save space, we use the simplified notations, such as
$~(\g\low b\chi)_\a \equiv (\g\low b)\du\a\b\chi_\b$, and $~(\g\ldu{b c
d}{e f} T_{e f})_\a \equiv  - (\g\ldu{b c d}{e f} )_{\a\b} T\du{e f}\b$, {\it
etc.},  while  the symbol
$~{\scst \[n\]}$~ is use for totally  antisymmetric bosonic ~$n$~ indices
{\it e.g.,} $~ (\g^{\[3\]})_{\a\b} \Tilde N_{\[3\]}
\equiv  (\g^{a b c})_{\a\b} \Tilde N_{a b c} $.   The$~\Tilde N$'s is the 
Hodge dual of $~N\low{\[7\]}$.  We use $~A_{\[3\]}$~ in (3.2j) as an   
important building block for coupling supergravity to the VM, because 
this $~A\-$tensor will accommodate higher-order 
superstring corrections \bffcdual.  

The superfield equations are obtained from the BIs at the 
engineering dimensions $~d\ge 3/2$, as  
$$  \li{ &  i (\nablasl \chi)^\a - \frac{2i} 9 (\g^{\[3\]} \chi )^\a
       \Tilde N_{\[3\]} 
     - \frac{4i} 3 (\g^a \chi )^\a \nabla_a\Phi 
\cr 
& \hskip 1.5in 
- \frac i 3 (\g^{\[3\]}\nabla)^\a A_{\[3\]} 
        - \frac{2{\sqrt2} i }9 (\g^{\[3\]} \chi)^\a  
      A_{\[3\]}  \eqdot 0 {~, ~~~~~ ~~~~~ ~~~ }  
&(3.3\rma) \cr  
\vbox{\vskip 0.3in} 
&  i (\g^b T_{a b} )_\g + \frac{2\sqrt2} 3 \nabla_a \chi\low\g
       + \frac 1{27\sqrt2} (\g\du a{\[3\]} \chi )_\g \Tilde N_{\[3\]} 
      - \frac 5 {9\sqrt2} (\g^{b c} \chi )_\g \Tilde N_{a b c} 
     - \frac {8{\sqrt2} i} 3 \chi\low\g \nabla_a \Phi   \cr 
&  ~~~~~ ~~~~~ ~~~~~    + \frac{\sqrt 2}{189} 
       ( \g^{b c} \chi)_\g \chi\low{a b c} 
       - \frac {20}{189} (\g\du a{\[3\]}\chi )_\g A_{\[3\]} 
       + \frac 4 9 (\g^{b c}\chi)_\g A_{a b c} \cr 
&  ~~~~~ ~~~~~ ~~~~~   - \frac{5\sqrt2}{63} 
      (\g\du a{\[3\]} \nabla)_\g A_{\[3\]} 
    + \frac{\sqrt2} 3 (\g^{b c}\nabla)_\g A_{a b c} 
    \eqdot 0 {~~,  ~~~~~ ~~~~~ } 
&(3.3\rmb) \cr 
\vbox{\vskip 0.3in} 
& (\g^{a b})\ud\g\d \,  T\du{a b}\d \eqdot 0 ~~, 
&(3.3\rmc) \cr 
\vbox{\vskip 0.3in} 
& R_{a b} - \frac 4 3 \nabla_a \nabla_b \Phi 
       + \frac {16} 9 (\nabla_a\Phi) (\nabla_b\Phi) 
       - \frac 19 \eta\low{a b} \Tilde N_{\[3\]}{}^2 
       + \frac 4 3 \Tilde N\du{a b} c \nabla_c \Phi \cr 
& ~  + \frac i 9 (\Bar\chi\g\low b \nabla_a \chi) 
       + \frac{\sqrt 2} 3 (\Bar\chi\g\low a {}^c T_{b c} ) 
       + \frac1{2\sqrt 2} (\Bar\chi \g\ldu b d T_{a d}) 
       - \frac 5{6\sqrt 2}  (\Bar\chi \, T_{a b})  \cr 
& ~ + \frac 1{324} \eta\low{a b} \chi^{\[3\]} \Tilde N_{\[3\]} 
      - \frac 1{108}\chi\low{a\[2\]} \Tilde N\du b{\[2\]}  
      - \frac 5{108}\chi\low{b\[2\]} \Tilde N\du a{\[2\]} 
        + \frac{5i}{252\sqrt2} 
        \eta\low{a b} (\Bar\nabla\g^{\[3\] }\nabla) A_{\[3\] }  \cr 
& ~  - \frac{5i}{84\sqrt2} (\Bar\nabla\g\ldu a{c d}\nabla)  
        A_{b  c d} 
        - \frac i {12\sqrt2} (\Bar\nabla\g\ldu b{c d}\nabla) A_{a c d} 
       + \frac{2\sqrt2} 3 \nabla_c A\du{a b} c 
       + \frac{10\sqrt2} {189} \eta\low{a b} \Tilde N^{\[3\]} A_{\[3\]} \cr 
& ~ - \frac{26\sqrt2}{21} \Tilde N\du a{c d} A_{b c d} 
      - \frac {38\sqrt2}{21} \Tilde N\du b{c d} A_{a c d} 
      - \frac{5\sqrt2}{567} \eta\low{a b} \chi\low{\[3\]} A^{\[3\]} 
        + \frac{5\sqrt2}{189} \chi\ldu a{c d} A_{b c d} 
        + \frac{\sqrt2}{27} \chi\ldu b{c d} A_{a c d} \cr 
& ~  - \frac{8\sqrt2}9 (\nabla_c\Phi) A\du{a b} c 
       - \frac{80}{63} \eta\low{a b} A_{\[3\]}^2 
       + \frac{64} 7 A\du a{c d} A_{b c d} 
      + \frac {5i} {378} (\Bar\chi\g\ldu{a b}{\[3\]}\nabla) A_{\[3\]} 
     + \frac{5i}{378} \eta\low{a b} (\Bar\chi \g^{\[3\]} \nabla) A_{\[3\]} \cr 
& ~ - \frac{5i}{126} (\Bar\chi \g\ldu a{c d} \nabla) A_{b c d} 
     - \frac i{18}(\Bar\chi\g\ldu b{c d} \nabla) A_{a c d} 
      + \frac i 9 (\Bar\chi \g^c\nabla) A_{a b c}      
\eqdot 0 { ~, ~~~~~ ~~~~~ }  
&(3.3\rmd) \cr 
\vbox{\vskip 0.3in} 
& R_{\[a b\]} \eqdot 0 ~~, 
~~~~ R - \frac 4 3 \Tilde N_{\[3\]}{}^2  \eqdot 0 ~~, 
&(3.3\rme) \cr 
\vbox{\vskip 0.3in} 
& \nabla_a^2 \Phi - \frac 16 \Tilde N_{\[3\]}^2 
      - \frac 43 (\nabla_a\Phi)^2 \cr 
& ~~~~~ ~ - \frac i{24\sqrt2} (\Bar\nabla\g^{\[3\]}\nabla) A_{\[3\]} 
      + \frac{17\sqrt2} 9 \Tilde N_{\[3\]} A^{\[3\]} 
       + \frac 8 3 A_{\[3\]}^2 
       - \frac i{18} (\Bar\chi\g^{\[3\]}\nabla)A_{\[3\]} 
      \eqdot 0 ~~,  
&(3.3\rmf) \cr 
\vbox{\vskip 0.3in} 
& \nabla_{\[a} \Tilde N_{b c d\] } - 3 \Tilde N\du{\[a b} e \Tilde N_{c d\] e}  
    + \frac 4 3 \Tilde N_{\[a b c} \nabla_{d\]} \Phi 
     - \frac 1{\sqrt 2} (\Bar\chi\g_{\[a b} T_{c d\]} ) 
     +  \frac i{2\sqrt2} (\Bar\nabla\g_{a b c d}\g^{\[3\]}\nabla)
      A_{\[3\]}  \cr 
& ~~~~~ - 12{\sqrt2} \nabla_{\[a} A_{b c d\]} 
      + 34{\sqrt 2} \Tilde N\du{\[a b} e A_{c d\] e} 
      + \frac{40\sqrt2} 3 (\nabla_{\[a} \Phi) A_{b c d\]} 
      + 48 A\du{\[a b} e A_{c d\] e} \cr 
& ~~~~~ + \frac i 3 (\Bar\chi \g\low{a b c d} \g^{\[3\]}\nabla)A_{\[3\]} 
       + \frac i 3 (\Bar\chi \g^{\[3\]} \g_{a b c d} \nabla) A_{\[3\]}   
       \eqdot 0 ~~,     
&(3.3\rmg) \cr 
\vbox{\vskip 0.3in} 
& i (\nablasl \l^I) - \frac i 6 (\g^{\[3\]} \l^I) \Tilde N_{\[3\]} \eqdot 0 ~~, 
&(3.3\rmh)  \cr 
\vbox{\vskip 0.3in} 
& \nabla_b F\du a {b\, I} + \frac i{\sqrt 2} (\Bar\l{}^I \g_b T\du a b) 
       \eqdot  \nabla_b F\du a {b\, I} 
       - \frac i{2\sqrt 2} (\Bar\l{}^I\g\du a {b c} T_{b c} ) \eqdot 0  ~~. 
&(3.3\rmi) \cr } $$ 
We use the symbol $~\eqdot$~ for a field equation, and 
$~(\Bar\nabla\g^{\[3\]} \nabla) \equiv (\g^{\[3\]})^{\a\b} 
\nabla_\b\nabla_\a$, {\it etc.}  The last two 
superfield equations (3.3h,i) are for the usual VM.    

As in the case of the usual BFFC in \gnz, all the exponential factors  with
the dilaton $~\Phi$~ have disappeared, making the whole system 
drastically simplified \bffcdual.  We also remind the readers that  this
BFFC \bffcdual\ is one of the simplest constraint sets among other
possible  sets connected by super-Weyl rescalings 
\ref\gv{S.J.~Gates, Jr.~and S.~Vashakitze, \np{291}{87}{172}.}. 
This leads to drastic simplification of the couplings of the dual VM 
in the next section.  For example, the duality relationship (3.3c) 
between the fields strengths of usual and dual VMs will be clear with no 
additional fermionic terms.     

Note also that it is only the dual version of $~N=1$~ supergravity 
\gnanomaly\gngsstring\bffcdual\ that can couple to the dual VM
consistently, as the presence of  the $~N\wedge F\-$term reveals.

\bigskip\bigskip\medskip 


\leftline{\bf 4.~~Dual VM Coupled to $~N=1$~ Supergravity}   

Armed with the BFFC for $~N=1$~ dual supergravity \bffcdual, we are 
now ready to consider the dual VM couplings.  The VM up to now can
carry non-Abelian adjoint index ${\scst I}$, but in this section, we
consider no additional index on the dual VM.  Our dual VM has the field
content $~(C_{m_1\cdots  m_7}, \l^\a)$, where $~\l^\a$~ is the same
gaugino as in the last section,  {\it i.e.,} a Majorana-Weyl spinor in 10D
with positive chirality.  In superspace, the 7-form potential
$~C_{A_1\cdots A_7}$~  has its 8-form superfield strength
$~H_{A_1\cdots A_8}$.  The spatial  component $~H_{a_1\cdots a_8}$~ is
dual to $~F_{a b}$, and the 7-form potential $~C_{a_1\cdots a_7}$~ has 
the same 8 physical degrees of freedom as $~A_a$.   

The most important key ingredient for the coupling of the dual 
VM to $~N=1$~ dual supergravity is the introduction of the 
$~N\wedge F\-$term in the $~H\-$BI:  
$$ \li{ & \frac1{8!} \nabla_{\[A_1} H_{A_2\cdots A_9)} 
      - \frac 1{7!\cdot 2} T\du{\[A_1 A_2|} B H_{B | A_3\cdots A_9)} 
        + \frac 1{7!} N_{\[A_1\cdots A_7} F_{A_8 A_9 ) } \equiv 0 ~~. 
&(4.1) \cr } $$ 
Note that the coefficient of the last term is $~1/7!$~ instead of 
the commonly expected one $~1/(7! \cdot 2)$.  This is due to 
our notation, and nothing else.  It is essential to consider 
this $~H\-$BI, in addition to the other BIs in (4.1) for our whole system.      

Even though the presence of the 2-form superfield strength $~F$~ in
(4.1), when dealing with the dual VM, is bizarre at first glance, this
situation is similar to the duality symmetric formulation of 11D
supergravity 
\ref\berkovitsetal{I.~Bandos, N.~Berkovitz  and D.~Sorokin,
\np{522}{98}{214}, \hepth{9711055};  H.~Nishino, \mpl{14}{99}{977},
hep-th/9802009.}.   
In the latter, both the usual 4-form superfield strength $~F_{A B C D}$~
and its dual  7-form superfield strength $~G_{A_1\cdots A_7}$~ are 
present at the same time.  To put it differently, we can maintain the
constraints for each constraints of $~F_{A B}$~ related to (3.2e) or (3.2f)
even in the formulation of dual VM.  This is nothing unusual in 
superspace, when we deal with dual systems \berkovitsetal.    

The necessity of the last $~N\wedge F\-$term in (4.1) is  understood as
follows.  If we did not have this term, then a problem occurs at
$~d=1/2$~ with the $~H\-$BI:  $~(1/2) \nabla_{(\a}  H_{\b\g)  d_1\cdots
d_6}  + \cdots\equiv 0$.  A $~\chi\-$linear 
term develops with no counter-term to cancel.  It is  exactly the combination
$~N_{(\a\b| \[ d_1\cdots d5} F_{d_6\] |\g)} $~ from the $~N\wedge
F\-$term that cancels this unwanted term.   A similar cancellation is
observed in the $~H\-$BI: $~\nabla_{(\a} H_{\b) c_1\cdots c_7}+
\cdots\equiv 0 $~ at $~d=1$.   The combination $~N_{\a\b
\[c_1\cdots c_5}  F_{c_6 c_7\] }$~ cancels the otherwise unwanted term
linear in $~\Tilde  H_{c_6 c_7}$,  yielding also the on-shell duality
relationship $~F_{a b} \eqdot \Tilde H_{a b}$~ as in (4.3c) below.   

We emphasize that the modification of the $~H\-$BI 
is the solution to the  problem of
coupling the dual VM to supergravity.  In particular, the 
structure of the $~N\wedge F\-$term is similar to the case of the  
duality-symmetric formulation of 11D supergravity 
\ref\elevendual{M.~Cederwall, B.E.W.~Nilsson and 
P.~Sundell, \jhep{9804}{98}{007}, \hepth{9712059};  
I.A.~Bandos, N.~Berkovits and D.P.~Sorokin, 
\np{522}{98}{214}, \hepth{9711055};  
H.~Nishino, \mpl{14}{99}{977}, \hepth{9802009}.},       
or the superspace formulation of the massive Type IIA supergravity 
in 10D 
\ref\supereight{H.~Nishino, \pl{457}{99}{51}, \hepth{9901027}.}.    
Our $~H\-$BI is analogous to \elevendual,  in the sense that the superfield
strength is involved in the $~H\-$BI.  However, it is more  similar to
\supereight, in the sense  that the superfield strength $~N$~ is 
multiplied by $~F$.  

The new constraints related to the dual VM at $~d\le 1$~ are 
$$ \li{ & H_{\a b_1\cdots c_7}    
       = - \frac i{\sqrt2} (\g\low{c_1\cdots c_7} \chi)_\a~~,
&(4.2\rma) \cr 
& \nabla_\a\l^\b \eqdot - \frac1{2\sqrt 2} (\g^{c d})\du\a\b 
       \Tilde H_{c d} ~~. 
&(4.2\rmb) \cr } $$ 
For the reasons already mentioned, and to be clarified shortly, 
the fact that the gaugino $~\l$~ appears both at (4.2b) and (4.2f) 
does not pose any problem.

The dual VM superfield equations are our new ingredients here.  
They can be obtained from our new $~H\-$BIs at $~d\ge 1$:  
$$ \li{ & i (\nablasl \l) - \frac i 6 (\g^{\[3\]} \l ) \Tilde N_{\[3\]} \eqdot 0 ~~, 
&(4.3\rma)  \cr 
& \nabla_b \Tilde H\du a b + \frac i{\sqrt 2} (\Bar\l \g^b T_{a b} ) 
       \eqdot  \nabla_b \Tilde H\du a b
       - \frac i{2\sqrt 2} (\Bar\l\g\low a{}^{b c} T_{b c} )  \eqdot 0 ~~, 
&(4.3\rmb) \cr 
& F_{a b} \eqdot \Tilde H_{a b} 
    \equiv \frac1{8!} \e\low{a b}{}^{\[8]} H_{\[8\]} ~~. 
&(4.3\rmc) \cr } $$ 
Eq.~(4.3a) is exactly the same as (4.3h), but is repeated here  because
$~\l$~ also belongs to the dual VM.   Compared with the ordinary case,
the on-shell duality  relationship (4.3c) comes out of $~H\-$BI at
$~d=1$.  This result  is common to supergravity theories involving
dualities.  Note that (4.3c) is  also associated with the cancellation
between the $~N\wedge F\-$term  and $~\Tilde H\-$term at $~d=1$~ as
mentioned before.  

Under (4.3c), we can replace $~F_{a b}$~ everywhere by $~\Tilde H_{\[8\]}$, 
and {\it vice versa} everywhere in our system.  For example, 
(4.2f) has $~F$~ on the r.h.s., but now it also implies (4.2b).  
This tells us that the $~\l\-$field belongs both to the original 
VM $(A_a, \l^\a)$~ and the dual VM $(C_{\[7\]}, \l^\a)$~ at the 
same time.  Another example is (4.2h) with $~F_{b c}$, 
now implying that 
$$\li{ \nabla_\a \Tilde H_{b c} 
\eqdot & - \frac i{\sqrt 2} (\g\low{\[b} \nabla_{c\]} \l)_\a 
     - \frac {5i}{3\sqrt 2} (\g^d\l)_\a \Tilde N_{b c d}   \cr 
& + \frac i{6\sqrt 2} (\g\low{\[b | }{}^{\[2\]} \l)_\a 
          \Tilde N_{| c\] \[2\]} - \frac i{18\sqrt2} (\g\low{b c}{}^{\[3\]} \l)_\a 
          \Tilde N_{\[3\]} ~~.   
&(4.4) \cr } $$ 
Accordingly, the couplings between the dual VM and $~N=1$~ 
dual supergravity are through the $~A_{\[3\]}\-$tensor, 
{\it i.e.,} the bilinear form in $~\l$~ exactly as in (4.2j), except that 
the $~\l$'s now carries no adjoint index $~{\scst I}$.    

There are several consistency checks to be performed.  A typical one is 
the $~H\-$BI at $~d=2$~ which yields immediately (4.3b).  In particular, 
a possible $~\Tilde N^{a c d} \Tilde H_{c d}\-$term in (4.3b) is cancelled by 
the peculiar $~N\wedge F\-$term in the $~H\-$BI.  This makes (4.3b) 
equivalent to  (4.3i) under $~\Tilde H\eqdot F$~ (4.3c).  We can also 
confirm the vanishing of the divergence of (4.3b), after using  the
relevant superfield equations.  

The existence of the $~N\wedge F\-$term in our $~H\-$BI implies 
the necessity of the Chern-Simons term in the ~$H\-$superfield 
strength itself:  
$$ \li{ & H_{A_1\cdots A_8} 
    \equiv \frac1{7!} \nabla_{\[A_1} C_{A_2\cdots A_8)} 
     - \frac1{6!\cdot 2} T\du{\[ A_1 A_2 |}B C_{B|A_3\cdots A_8)} 
     - \frac1{6!} M_{\[A_1\cdots A_6} F_{A_7 A_8)} 
     {~~. ~~~} 
&(4.5) \cr } $$ 
This implies, in particular, that the bosonic field strength  with 
the Chern-Simons term 
$$ \li{ & H_{m_1\cdots m_8} 
      \equiv \frac1{7!} \partial_{\[m_1} C_{m_2\cdots m_8\]} 
     - \frac1 {6!} M_{\[m_1\cdots m_6} F_{m_7 m_8\]} ~~,  
&(4.6) \cr } $$ 
which is a new result.  

Notice that the `duality relationship' exists 
not only between the component $~F_{a b}$~ and $~H_{\[8\]}$, 
but also between $~F_{\a b}$~ and $~H_{\a \[7\]}$, as can be 
verified by (4.2e) and (4.2a):    
$$\li{ & F_{\a b} = \frac 1{9!} \e\ldu b {c d \[7\]}
      (\g_{c d})\du\a\b H_{\b \[7\]} ~~. 
&(4.7) \cr } $$   
This suggests the existence of the superspace duality  
$$ \li{ & F_{A B} \eqdot  \frac1{8!} \E \du{A B}{C_8\cdots C_1} 
     H_{C_1\cdots C_8} ~~,  
&(4.8) \cr} $$ 
with the generalized $`\e\-$tensor' defined by 
$$\li{ & \E \du{A B}{C_1\cdots C_8} 
      \equiv \cases{ \E\du{a b}{c_1\cdots c_8}
      \equiv \e\ldu{a b}{c_1\cdots c_8} ~~, \cr 
\vbox{\vskip 0.2in} 
      \E\du{\a b} {\g c_1\cdots c_7} 
      \equiv \frac19\e\ldu b{d e c_1\cdots c_7} (\g_{d e} )\du\a\g ~~, \cr
\vbox{\vskip 0.2in} 
      \E \du{\a \b} {\g_1 \g_2 c_1\cdots c_6} 
     \equiv \frac1{10} \e^{d e f g c_1\cdots c_6} (\g\low{d e} )\du\a{\g_1}
     (\g\low{f g})\du\b{\g_2}  ~~.  \cr }  
&(4.9) \cr } $$
In fact, a similar generalized duality relationship in 10D was found 
\gngsstring\ between the 3-form superfield strength 
$~G_{A B C}$~ in usual supergravity \chamseddineusual\ 
and the 7-form $~N_{A_1\cdots A_7}$~ 
in dual supergravity \gnanomaly\gngsstring.  

Some readers may wonder, whether it makes sense to have  both the
usual VM superfield strength $~F$~ and  the dual VM superfield strength
$~H$~ simultaneously.   It looks bizarre at first glance, because it seems
to contradict  the counting of the physical degrees of freedom. 
However, this puzzle can be easily solved by the following points.   First,
note that there is a on-shell duality relation  
$~F\eqdot \Tilde H$ (4.3c).  Therefore, even if we temporarily double the
number of degrees of freedom from $~8$~ of the usual vector
$~A$~ into $~8+8 = 16$~ for $~A$~ and $~C$, the duality
relationship $~F\eqdot \Tilde H$~ reduces it to $~16/2 = 8$.   This
situation is not new, but similar to the case of duality symmetric
formulation for 11D supergravity \elevendual.   The simultaneous
relationships (4.2f) and (4.2b) are also consistent  with this counting,
namely, the gaugino field $~\l$~ belongs  both to the original VM and
the dual VM simultaneously.  

We mention an alternative answer to the question of  degrees of
freedom, based on  the superspace duality (4.8).  Namely, since all the
$~F_{A B}$~  can be formally expressed in terms of $~H_{A_1\cdots A_8}$, 
there is no need to  introduce $~F_{A B}$~ itself and $~F\-$BIs 
themselves (4.1c), either.  In other words, once we replace $~F_{A B}$~ 
everywhere by $~H_{A_1\cdots A_8}$~ under (4.8), 
we can totally forget about the existence of $~F_{A B}$, and therefore,  
of the usual VM.  With this prescription, even the $~N\wedge F\-$term 
will contain only the $~H\-$superfield strength:
$$ \li{ & \frac1{8!} \nabla_{\[A_1} H_{A_2\cdots A_9)} 
        - \frac 1{7!\cdot 2} T\du{\[A_1 A_2|} B H_{B | A_3\cdots A_9)} 
        + \frac 1{7!\cdot 8!} N_{\[A_1\cdots A_7} 
       \E\du{A_8 A_9 ) }{B_1\cdots B_8} H_{B_1\cdots B_8} \equiv 0 
       {~. \hskip 0.6in}  
&(4.10) \cr } $$  

Accordingly, the field strength $~F~$ in the Chern-Simons term
$~M\wedge F$~  in (4.5) is replaced by $~H$~ itself.  This  reflects nothing
but the non-linear nature of the  duality transformation \nt\ for the dual
VM in component  language.  In fact, in the dual version of supergravity
\gnanomaly, the
$~M\wedge F \wedge F\-$term in the lagrangian already suggests such
a non-linearity for the duality transformation $~F\rightarrow H$. 
Relevantly, such a duality transformation is possible only in the  dual
version \gnanomaly, but not in the usual version \chamseddineusual\ of
supergravity.  This is because the VM coupling in the latter is in the
Chern-Simons term
$~F\wedge A$~ in the $~G\-$field strength with the `bare' potential
$~A$~ preventing a duality transformation \nt.  On the other hand, in 
the dual version \gnanomaly, the corresponding term 
$~M\wedge F\wedge F$~ is only in
terms of the field strength $~F$, enabling   
the duality transformation $~F\rightarrow H$~ in a non-linear 
fashion.

\newpage

\leftline{\bf 5.~~Anomaly Cancellation} 

A pressing question to ask is whether our model makes sense at 
the quantum level, because it lacks a more fundamental  
formulation such as superstring theory \gsw.  In fact,  
this question can be rather easily answered by 
considering the following points.  First, our gauge group is Abelian 
with {\it no} minimal couplings with fermions.  As such,  
the purely gauge and mixed anomalies are absent.  

As for purely gravitational anomalies, notice that 
the anomaly 12-form $~I_{12}$~ is given by \gs 
$$ \li{ I_{12} =\, & \Big( \fracmm{n-496}{7560} \Big) \, \left[ \,  \tr R^6  
     + \fracm{21}{16} \tr R^2 \tr R^4 
     + \fracm{35}{64} (\tr R^2)^3 \right] \cr 
& + (\tr R^2) \left[ \,  \frac1{32} (\tr R^2)^2 
      + \frac 18 \tr R^4 \, \right]  { ~~.  ~~~~~ }
&(5.1) \cr } $$ 
Compared with \gs, due to the absence of pure gauge and mixed
anomalies, no terms containing the gauge field strength are 
involved.  The number $~n$~ is the number of the $~C\-$fields.    
When $~n=496$, the leading term $\,\tr R^6\,$ vanishes, 
while all the non-leading terms factorize as
$$\li{ & I_{12} = (\tr R^2) X_8 ~~, \cr 
& X_8 \equiv \frac 1 8 \tr R^4 + \frac 1{32} (\tr R^2)^2 ~~. 
&(5.2) \cr } $$
The consistent anomaly corresponding to $~I_{12}$~ is 
\ref\bz{W.A.~Bardeen and B.~Zumino, \np{244}{83}{421}.}
$$ \li{ & \G= \int \left[ \, 2 \, \o^1_{2\rm L} X_8 
     + 4 (\tr R^2) X_6^1 \, \right] ~~,  
&(5.3) \cr } $$ 
where the $~\o$'s and $~X$'s are defined by 
$$ \li{ & d \, \o\low{3\rm L} = \tr R^2 ~~, ~~~~ 
    \d\, \o\low{3\rm L} = - d \, \o^1_{2\rm L} ~~, \cr 
& X_8 = d X_7 ~~, ~~~~\d X_7 = - d X_6^1 ~~,  
&(5.4) \cr } $$
under a local Lorentz transformation $~\d$.  
Now consider the counter-term 
$$ \li{ & S_c \equiv \int \left[ -6 M \, (\tr R^2) 
           - 2 \, \o_{3\rm L} X_7 \, \right]  ~~, 
&(5.5) \cr } $$ 
where $~M$~ is the 6-form tensor present in the 
dual supergravity multiplet \gnanomaly\gngsstring.  
The anomaly $~\G$~ is cancelled by the variation 
$$ \li{ & \d M = X_6^1 ~~,    
&(5.6) \cr } $$ 
because 
$$ \li{ & \d I (N\rightarrow \Tilde N) + \d S_c + \G = 0 ~~, 
&(5.7) \cr } $$ 
for the the original action $~I$, and 
~$\Tilde N$~ is the modified field strength of $~M$:   
$$ \li{ & \Tilde N \equiv dM + X_7~~, ~~~~\d \Tilde N = 0 ~~, ~~~~
     d \Tilde N = X_8~~. 
&(5.8) \cr }  $$ 

Thus we need ~496~ copies of the dual VMs for  
$~U(1)^{496}$, where each multiplet has couplings to  supergravity. 
Our total lagrangian is $~\Lag_{\, \rm total}  = \sum_{i=1}^{496}
\Lag_i$, where $~\Lag_i$~ has exactly the same coupling structure of 
section 3 for the $~{\scst i}\-$th multiplet $~(C_{m_1\cdots m_7}{}^i ,
\l^i)$.  Note that these  496 dual VMs form just a direct sum, with no
mutual interaction  among themselves.    

The anomaly freedom indicates that despite the lack of a more
fundamental theory such as superstring theory \gsw, the dual VM still
has its proper {\it raison d'etre} as a consistent interacting theory in 10D,
at least, within a point field theory formulation.  It is imperative that
dimensional reduction of our theory to lower-dimensions will generate 
a series of anomaly-free  theories in those dimensions.  

\bigskip\bigskip\medskip 

\leftline{\bf 6.~~Supersymmetric Dirac-Born-Infeld Interactions 
for dual VM}   

We give here our result for an invariant action for the 
Abelian dual VM with globally supersymmetric DBI interactions \dbi%
\ref\mr{R.R.~Metsaev and M.A.~Rahmanov, 
\pl{193}{87}{202}.}%
\ref\brs{E.~Bergshoeff, M.~Rakowski and E.~Sezgin, 
\pl{185}{87}{371}.}%
\cnt.
Our action $~I_{\rm SDBI} \equiv \int d^{10} x\, \Lag_{\rm SDBI}$~ 
is given by the lagrangian
$$\li{ \Lag_{\rm SDBI}  
= & - \frac 1{2 (8!) }  (H_{\[8 \] })^2 - \frac i 2 (\Bar\l \g^a \partial_a  \l) 
    - \frac 1 4 \a^2  (\Tilde H^4)\du a a
     + \frac 1{16} \a^2 [(\Tilde H^2) \du a a]^2 \cr 
&  - \frac i 2 \a^2  (\Tilde H^2)^{a b} (\Bar\l\g_a\partial_b \l) 
     + \frac i 8 \a^2 \Tilde H\du a d (\partial_d\Tilde H_{b c} ) 
     (\Bar\l\g^{a b c } \l) 
     + \frac 1{12} \a^2  (\Bar\l\g_a \partial_b \l)^2 
      ~~. \hskip 0.35in 
&(6.1) \cr } $$ 
The multiplication of the $~\Tilde H$'s is defined, {\it e.g.,} by 
$~(\Tilde H^2)\du a b \equiv 
\Tilde H\du a c\Tilde H\du  c b$, {\it etc.}     
The third and fourth terms in the first line are the 
DBI-interactions \dbi\ at $~\calO(\a^2)$, in terms of our dual field 
strength $~\Tilde H$, where the constant $~\a$~ has 
the engineering dimension of $~(\hbox{length}) = (\hbox{mass})^{-1}$.  

Our action $~I_{\rm SDBI}$~ is invariant up to $~\calO (\a^3)$~ 
under supersymmetry 
$$ \li{ \d_Q C_{\[7\]} 
      = \, & +\frac i{\sqrt2} (\Bar\e\g\low{\[7\]} \l) \cr 
& \hskip -0.53in + \frac 1{6\sqrt2} \a^2 \e\ldu{\[7\]}{b c d} 
         \Big[  - \frac i 8 (\Bar\e\g\low{b c d} \l) (\Tilde H^2)\du e e 
      - 3 i (\Bar\e\g^e\l) \Tilde H_{e b} \Tilde H_{c d} \cr 
& \hskip 0.65in - \frac {3i} 2 (\Bar\e\g\low{e b c} \l) (\Tilde H^2)_{e d} 
      + \frac i{16} (\Bar\e \g\ldu{b c d}{e f g h}\l) \Tilde H_{e f} 
     \Tilde H_{g h} \cr 
& \hskip 0.65in  + \frac {3i} 4 (\Bar\e\g\ldu b{e f} \l) 
     \Tilde H_{e f} \Tilde H_{c d} 
     - \frac {3i} 2 (\Bar\e\g\ldu b{e f} \l) \Tilde H_{c e} \Tilde H_{d f} 
      \, \Big] + \calO(\a^2\l^2) {~~, ~~~~~ ~~~~~} 
&(6.2\rma) \cr
\d_Q\l \, = & - \frac 1{2\sqrt2} ( \g^{a b} \e) \,  \Tilde H_{a b} 
       - \frac 3 {16\sqrt2} \a^2  ( \g^{c d} \e ) \,  
       \Tilde H_{c d} (\Tilde H^2) \du e e   
      - \frac 3{4\sqrt2} \a^2  ( \g^{a b} \e ) \,  (\Tilde H^3)_{a b} \cr
& + \frac 1{96\sqrt2}  \a^2  ( \g^{a b c d e f} \e) \,  
     \Tilde H_{a b} \Tilde H_{c d} \Tilde H_{e f}  
    + \calO( \a^2 \l\Tilde H ) ~~.  
&(6.2\rmb) \cr } $$  

The field equations for the $~C$~ and $~\l\-$fields are 
$$ \li{  \fracmm{\d}{\d C_{\[7\] }} \Lag_{\rm SDBI} 
= &-\frac1{12} \e^{\[7\] a b c} 
     \partial_a \Big[ \, \Tilde H_{b c} 
      - 2 \a^2 (\Tilde H^3)_{b c} 
      + \frac 12 \a^2(\Tilde H^2)\du e e \Tilde H_{b c} 
      - i \a^2  \Tilde H\ud e b (\Bar\l\g_{(e }\partial_{c) } \l)  \cr 
& \hskip .9 in + \frac i 4 \a^2 (\partial_b \Tilde H_{e f} ) 
       (\Bar\l \g\ldu c{e f} \l) 
      + \frac i 2 \a^2 \Tilde H^{e f} (\Bar\l\g\low{b c e} \partial_f \l) \,
        \Big] \eqdot 0 { ~, \hskip 0.5in}  
&(6.3\rma) \cr 
\fracmm{\d}{\d \l} \Lag_{\rm SDBI} 
= & - i  \delsl \l 
    - i \a^2 (\g_a \partial_b \l) (\Tilde H^2)^{a b} 
    - \frac i2 \a^2 (\g_a \l)\partial_b (\Tilde H^2)^{a b} \cr 
&  + \frac i 4 \a^2  (\g^{a b c} \l )
     \Tilde H\du a d \partial_d \Tilde H_{b c}  
    + \frac 1 3 \a^2  (\g_a \partial_b \l)
      (\Bar\l\g^a \partial^b \l) \eqdot 0 ~~.    
&(6.3\rmb) \cr } $$
The leading term in the 
$~C\-$field equation is like $~\partial_{\[a}  \Tilde H_{b c]} + \cdots$~ 
which is dual to the Bianchi identity $~\partial_{\[a } F_{b c\]}$~ for 
the original $~F\-$field strength.  On the other hand, the 
original $~A\-$field equation $~\partial_a F^{a b} + \cdots \eqdot 
0$~ becomes now an identity $~\partial_a \Tilde H^{a b}\equiv 0$.  

Our lagrangian  can be obtained by a direct construction, as we
can perform for the  conventional VM $(A_a, \l)$, with no  essential
obstruction, despite the subtlety about the extra symmetry.    
The extra symmetry at the free-field level arises in the 
commutator $~\[ \d_Q(\e_1) , \d_Q(\e_2) \] C_{a_1\cdots a_7}$~ as  
$$ \li{ & \d_{\rm E} C_{a_1\cdots a_7} 
      = \frac1{7!} \z\ldu{\[ a_1 a_2}{\[3\]} H_{a_3\cdots a_7\] \[3\]} ~~, 
&(6.4) \cr} $$ 
with the totally antisymmetric constant parameter $~\z^{\[5\]}$, 
leaving the $~H\-$field strength invariant up to the $~C\-$field 
equation:  
$$ \li{ & \d_{\rm E} \Tilde H_{a b} = \frac16 \z\ldu{a b}{c d e} 
\partial_{\[c } \Tilde H_{d e\]} \eqdot 0~~, 
&(6.5) \cr } $$ 
before interactions are switched on.  
In the past, it looked very difficult to maintain this extra symmetry, 
once interactions are switched on.   
However, our system seems to have overcome this 
problem, because the supersymmetry 
commutator shows that (6.5) is modified to  
$$\li{ \d_{\rm E} & \Tilde H_{a b} = \frac16 \z\du{a b}{c d e} 
        \partial_{\[ c | } \Big[ \Tilde H_{|d e\]} 
         - 2 \a^2 (\Tilde H^3)_{|d e\]} 
        + \frac 12 \a^2(\Tilde H^2)\du e e \Tilde H_{|d e\] } 
        - i \a^2  \Tilde H_{f |d|} 
       (\Bar\l\g^f \partial_{|e\]} \l)  \cr 
& \hskip - 0.1in - i \a^2  \Tilde H\ud f {|d|} 
        (\Bar\l\g\low{|e\]}\partial_f \l)  
         +  \frac i 4 \a^2 (\partial_{|d|} \Tilde H_{f g} ) 
         (\Bar\l \g\ldu{| e\]} {f g} \l) 
         + \frac i 2 \a^2 \Tilde H^{f g} (\Bar\l\g\low{|d e\] f} \partial_g\l) \,
          \Big] \eqdot 0 { ~~. \hskip 0.5in}   
&(6.6) \cr } $$  
This vanishes on-shell thanks to (6.3a).  
As is easily seen, the original free-field level relation (6.5) is now 
modified to include the supersymmetric DBI interactions, 
and it still vanishes upon using the $~C\-$field equation.  

Our lagrangian has a structure very similar to the conventional VM in 
\mr\brs.  Namely, the difference in coefficients occurs only for the 
 $~H^2 \l^2\-$terms, but not in the purely bosonic 
DBI-interactions or the quartic fermion term.  

In the actual derivation of our lagrangian, we have used the duality
transformation \nt\ from the conventional field strength $~F_{a b}$~ 
into the dual one $~H_{\[ 8 \]}$.  However, our system exhibits subtlety 
compared with the usual case.  In the usual duality transformation \nt\  
performed in supergravity in $~D\-$dimensions, 
we replace the original $~n\-$form 
$~F~$~ by a new independent $~n\-$form field 
$~G$, so that the kinetic term of the $~F\-$field  strength supplies
the bilinear in $~G$,  while the interactions are rather simple, such as 
the Pauli or the Noether terms always linear in $~G$.   The
constraint term $~\Lag_{\rm C} = C \wedge d G$~ \nt\ is 
also linear in $~G$, supplying the  desirable 
condition $~d G \eqdot 0$.  Therefore, in the usual case,  the duality
transformation is easy, because the $~G\-$field equation is  simply
algebraic $~G = \Tilde H + J $~ with a $~(D-n)\-$form $~H = d C$~ and 
an $~n\-$form source $~J$.   
If we try to apply this to the conventional VM with DBI interactions,  
the source $~J$~ now contains trilinear terms in $~G$~ due to the  
DBI interactions.  Therefore we can no longer solve the $~G\-$field 
equation $~G = \Tilde H + \calO(G^3)$~ algebraically for $~H$, 
and this was the  major obstruction in the past.   
However, in the present case, this problem does not arise, 
because the action for (6.1) is invariant up to $~\calO(\a^3)\-$terms.

\bigskip\bigskip\medskip 


\leftline{\bf 7.~~Concluding Remarks} 

In this paper, we have shown how to couple the dual VM 
$~(C_{m_1\cdots m_7}, \l)$~ to the dual $~N=1$~ supergravity 
\chamseddinedual\gnanomaly\gngsstring\bffcdual\ in superspace in
10D.    We have found that the peculiar new term 
$~N\wedge F$~ in the $~H\-$BI is the key ingredient for such couplings.  
Accordingly, the $~H\-$field strength should be modified by  the
Chern-Simons term $~M\wedge F$.  

Interestingly enough, our model is also free of anomalies for  the
particular group $~U(1)^{496}$.  In our theory,  no gauge or mixed
anomalies arise, while the purely gravitational anomaly has a vanishing
leading term for $~U(1)^{496}$, and all the non-leading terms are
cancelled by the variation of the 6-form tensor $~M$~ in the
supergravity multiplet by the Green-Schwarz mechanism 
\gs\gsw\gnanomaly\gngsstring.  Anomaly freedom is generally 
independent of the existence of superstring theory, but can be
formulated within point-field theory.  The anomaly freedom of our
theory strongly suggests the deep significance of such interactions, and
may lead to a more fundamental theory of extended objects which 
are not necessarily superstrings.  

Subsequently, we have also given the component lagrangian for the 
dual VM with consistent supersymmetric DBI interactions 
\dbi\mr\brs\cnt.  We presented our lagrangian including the fermionic
quartic terms.  We have shown the parallel structures between the 
conventional VM \mr\ and the dual VM at the lagrangian as well as  the
transformation levels.  The traditional problem with extra symmetry  in
the commutator algebra in component language has now been solved
by the $~\calO(\a^2)\-$modification of the extra symmetry  itself
consistent with the $~C\-$field equation.  
      
It is to be stressed that the presence of the $~N\wedge F\-$term  is very
crucial for the dual VM to couple to supergravity.  In particular, it is the
{\it dual} supergravity \gnanomaly\gngsstring\bffcdual\  that can
consistently couple to the dual VM.  This is a solution to the
long-standing problem with formulating higher-rank tensors in
superspace, in particular, with extra symmetries.  The necessity of such
a high-rank superfield strength as 
$~N$~ in superspace is also analogous to the  superspace formulation of 
the massive Type IIA supergravity in 10D \supereight.  

Once we have established dual VM interactions in 10D, we expect that
dimensional reductions will generate descendant anomaly-free
theories in lower-dimensions $~D\le 9$.  Our Abelian DBI interactions for
a dual VM may have interesting applications associated with non-linear
supersymmetry with Nambu-Goldstone mechanism 
\ref\bg{J. Bagger and A. Galperin, 
\pr{55}{97}{1091}, \hepth{9608177}.}.       

We emphasize that our formulation here is the first one giving  
non-trivial interactions between the dual VM and dual supergravity 
\chamseddinedual\gnanomaly\gngsstring\bffcdual\ in 10D, as well as
the Born-Infeld type self-interactions \dbi\mr\brs\cnt\  of the dual  VM. 
Even though our dual VM has only Abelian symmetry, our results must be
very crucial for future studies of dual VM in 10D.  In fact, the anomaly
freedom of our theory also suggests the existence of a more
fundamental theory of extended objects, which is not necessarily
superstring theory.

\bigskip

This work is supported in part by NSF Grant \# 0308246.


\vfill\eject 

\immediate\closeout\rfile\writestoppt
\baselineskip=14pt\centerline{{\bf References}}%
\bigskip{\frenchspacing%
\parindent=20pt\escapechar=` \input refs.tmp\vfill\eject}%
\nonfrenchspacing

\vfill\eject

\end{document}